\documentclass[12pt,onecolumn]{IEEEtran}
\usepackage{amssymb}
\usepackage{amsbsy}
\usepackage{amsmath}
\usepackage{graphicx}
\usepackage{amsfonts}
\newcommand{\mH}{\hbox{{\bf H}}}

\newcommand{\ga}{\alpha}
\newcommand{\gb}{\beta}
\newcommand{\grg}{\gamma}

\newcommand{\gre}{\varepsilon}

\newcommand{\gl}{\lambda}

\newcommand{\gs}{\sigma}

\def\bm#1{\mbox{\boldmath $#1$}}
\newcommand{\vga}{\mbox{$\bm \alpha$}}

\newcommand{\vgd}{\mbox{$\bm \delta$}}

\newcommand{\vgl}{\mbox{$\bm \lambda$}}
\newcommand{\vgm}{\mbox{$\bm \mu$}}

\newtheorem{theorem}{Theorem}[section]
\newtheorem{lemma}[theorem]{Lemma}

\newtheorem{prop}{Proposition}[section]
\newtheorem{claim}{Claim}[section]

\newtheorem{definition}{Definition}[section]
\newtheorem{question}{Question}[section]
\newtheorem{coro}{Corollary}[section]

\newcommand{\beq}{\begin{equation}}
\newcommand{\eeq}{\end{equation}}
\newcommand{\bea}{\begin{array}}
\newcommand{\ena}{\end{array}}
\newcommand{\bds}{\begin {itemize}}
\newcommand{\eds}{\end {itemize}}
\newcommand{\bdf}{\begin{definition}}
\newcommand{\blm}{\begin{lemma}}
\newcommand{\edf}{\end{definition}}
\newcommand{\elm}{\end{lemma}}
\newcommand{\bthm}{\begin{theorem}}
\newcommand{\ethm}{\end{theorem}}
\newcommand{\bprp}{\begin{prop}}
\newcommand{\eprp}{\end{prop}}
\newcommand{\bcl}{\begin{claim}}
\newcommand{\ecl}{\end{claim}}
\newcommand{\bcr}{\begin{coro}}
\newcommand{\ecr}{\end{coro}}
\newcommand{\bquest}{\begin{question}}
\newcommand{\equest}{\end{question}}

\usepackage{epsfig}
\pssilent

\title{Weighted Max-Min Resource Allocation for Frequency Selective Channels}
\author{Ephraim Zehavi$^1$, Amir Leshem$^1$, Ronny Levanda$^1$, Zhu Han$^2$
\thanks{Authors are with the $^1$School of Engineering, Bar-Ilan
University, Ramat-Gan, 52900, Israel and the $^2$ University of
Houston, USA. e-mail: leshema@eng.biu.ac.il. }}

\begin{document}
\bibliographystyle{ieeetr}
\maketitle

\begin{abstract}
In this paper, we discuss the computation of  weighted max-min rate
allocation using joint TDM/FDM strategies under a PSD mask
constraint. We show that the weighted max-min solution allocates the
rates according to a predetermined rate ratio defined by the
weights, a fact that is very valuable for telecommunication service
providers. Furthermore, we show that the problem can be efficiently
solved using linear programming. We also discuss the resource
allocation problem in the mixed services  scenario where certain
users have a required rate, while the others have flexible rate
requirements. The solution is relevant to many communication systems
that are limited by a power spectral density mask constraint such as
WiMax, Wi-Fi and UWB.

\end{abstract}

\section{Introduction}                  \label{sec:intro}
Orthogonal Frequency Division Multiple Access (OFDMA) is becoming a
ubiquitous technique for wireless multiple access schemes in
communication systems such as UWB, WLAN,  WiMAX and LTE, due to its
high spectral efficiency. OFDMA waveforms provide the flexibility of
allocating subcarriers to combat frequency selective fading. These
standards operate under two types of power constraints: Total power
and power mask; i.e. the Power Spectral Density (PSD) of the
transmitter is limited by the regulator. The total capacity of OFDMA
can be optimized by dynamically allocating subcarriers among users
according to channel conditions. However, the operator must satisfy
the subscribers' demands to provide a reasonable level of Quality of
Service (QOS). The standards define several different services that
allow QOS differentiation. The major challenges facing QOS in
wireless networks are the dynamic of the channels,  bandwidth
allocation, and handoff support. It is important to guarantee QOS at
each layer so that the network stays flexible. Bandwidth and bit
rates play a major role. They should be allocated in an efficient
manner. In some systems data services  and voice services have to be
supported simultaneously. These services  can conflict because voice
services are very delay sensitive and require real-time service.
Whereas, data services are less delay sensitive but are very
sensitive to loss of data and require almost-error-free
transmission. Thus both factors must be taken into account when
providing QOS for voice and data services. In this paper, we address
the allocation of subcarriers using a the weighted max-min approach
that sets user priority  according to a preset weight. This approach
is then  extended to guarantee a minimum data rate for voice
services and allocate the residue capacity to data services.

In \cite{jang03}a power adaptation method was suggested to maximize
users' total data rate in downlinks of an OFDM system. The
transmitted power adaptation scheme was derived by solving the
maximization problem in two steps involving subcarrier assignment of
users and power allocation of subcarriers. The outcome is that the
data rate of a multiuser OFDM system is maximized when each
subcarrier is assigned to only one user with the best channel gain
for that subcarrier, and the transmit power is distributed over the
subcarriers by a water-filling policy. However, fairness does not
enter into this approach. In the extreme case most of the spectrum
will be allocated to  a small group of subscribers with high average
channel gains.  In \cite{iyengar2006} the problem of resource
allocation of the OFDMA system was addressed. A heuristic scheduling
algorithm was proposed under the constraint that each subscriber
must obtain a preset data rate.

Rhee and Cioffi \cite{rhee2000} derived a multiuser convex
optimization problem under the total power constraint to find max-min  suboptimal subcarrier
allocation, where equal power is allocated to the subcarriers. A
max-min rate allocation algorithm maximizes the data rate of the
worst user, such that all users operate at a similar data rate.
However, this solution is not suitable when the operator has to
provide different level of services. Shen et al. \cite{shen05}
proposed a suboptimal proportional fairness resource sharing
mechanism which provides multiple service levels under total power
constraint while maximizing the total data rate. The algorithm
involves two steps. First, the subcarriers are allocated under the
assumption that the power is equal on each subcarrier. In the second
step, the power is distributed among the allocated subcarriers to
maximize the total rate while maintaining proportional fairness
constraints.
An alternative approach to the resource allocation problem is using game theoretic solutions such as the Nash bargaining solution under total power constraint  (see e.g., \cite{han2005}, \cite{nokleby07}), \cite{leshem2009gamenets} or under PSD mask constraint \cite{leshem2008} as well as the Kalai-Smorodinski solution \cite{park07}, \cite{zehavi09b},\cite{chen09}.

Here, we focus on power spectral density maks constraint and introduce the mechanisms to enable explicit subcarrier
allocation for multiple users in wireless systems when the following
conditions must be fulfilled:
\begin{enumerate}
\item  Differentiated service levels must be supported. A wireless operator should have the flexibility to specify differentiated service levels (or weights). The available radio
resource has to be partitioned proportionally to the weights.
\item  Voice service is  supported using a fixed  data rate.
\item  Computational
and signaling overhead must be minimal. A primary design goal of an
efficient resource allocation algorithm is to minimize the
communication and the computational load of feedback iterations
Algorithms have to be designed to calculate the allocation that puts
a minimal load on the system. Specifically, the time it takes to
calculate the fair rate must be minimal.
\end{enumerate}

In this paper, we show how the weighted max-min fairness design
criterion can assist operators in network optimization, at multiple
target rates. Here, we use a  model similar to \cite{shen05} but
employ a power mask rather than an average power constraint. It is
well known that  the total data throughput of a zero-margin system
is close to capacity even with a flat transmit (PSD) as long as the
energy is poured only into subcarriers with high SNR gains. A good
algorithm will not assign power to bad subcarriers. Furthermore, a
flat PSD might be necessary if the PSD mask constraint is tighter
than the total power constraint.

The remainder of the paper is organized as follows. In section II we
describe the general model of the wireless system and derive a
solution for the weighted max-min resource allocation problem.
Section III is focused on the special solution for the case of two
subscribers and outlines a simple algorithm for computing the
weighted max-min solution. Simulation results are presented and
discussed in Section IV.  Section V concludes this paper.

\section{Resource allocation using the weighted max-min solution}
In this section, we show that under a PSD mask constraint the
max-min fair solution can be computed using linear programming. This
is simpler than the total power constraint where general convex
programming is necessary. Assume that we have $N$ users, sharing a
frequency selective channel. Let the $K$ channel
matrices\footnote{These can be the uplink, downlink or multiple
source-destination pairs within the network.} at frequencies
$k=1,...,K$ be given by $\left<\mH_k:k=1,...,K\right>$. Each user is
allowed to transmit using a maximal power $p\left(k\right)$ in the
$k$'th subcarrier. In this paper, we limit ourselves to a joint FDM
and TDM scheme where an assignment of disjoint portions of the
frequency band to the various transmitters can be different at each
time instance as is done in Wimax. In the FDM/TDM case we have the
following:
\begin{itemize}
\item[1.] User $n$ transmits using a PSD limited by $\left<p_n(k): \ k=1,...,K \right>$.
\item[2.] Each user $n$ is allocated a relative time vector
$\vga=[\ga_{n1},...,\ga_{nK}]^T$ where $\ga_k$ is the proportion of time allocated to user $n$ at the $k$'th frequency channel. This is the TDM/FDM part of the scheme.
\item[3.] For each $k$, $\sum_{n=1}^N \ga_{nk}=1$. This is a Pareto-optimality requirement.
\item[4.] The rate obtained by user $n$ is given by
\begin{equation}
\begin{array}{c}
  R_n(\vga_n)=\sum_{k=1}^K \ga_{nk} R_{n k},
\end{array}.
\end{equation}
\end{itemize}
where,
\[
R_{n k} = \log_2 \left(1+\frac{|h_{nn}(k)|^2 p_n(k)}{\gs_n^2(k)}\right)
\]
and the subcarrier bandwidth is normalized to 1.
Interference is avoided by time sharing at each
frequency band; i.e, only a single user transmits at a given
frequency bin at any time. Furthermore, since at each time instance
each frequency is used by a single user, each user will transmit
using the maximal power.
{\em Note that we can replace the instantaneous rates by the long term averages using well known coding theorems for fading channels
\cite{biglieri98}. This allows much slower information exchange and makes the proposed approach practical in real  wireless systems}.

 The weighted max-min fair solution with weights
$\grg_1,...,\grg_N$ is given by solving the following equation: \beq
R_{\max \min}=\max_{\vga_1,...,\vga_N} \min_{1 \le n \le N}
\grg_nR_n(\vga_n). \eeq To solve this equation we rephrase it as a
linear programming problem: Let $c$ be the value of the weighted
max-min rate. We would like to maximize $c$ under the constraints
$R_n \geq c$, for all $1 \le n \le N$. Since each $R_n$ depends
linearly on $\vga_n$ we require \beq \max_{\vga_1,...,\vga_N,c}  c,
\eeq under the constraints
 \beq
 \bea{ll}
 0 \le c, \\
\frac{c}{\grg_n} \leq \sum_{k=1}^K \alpha_{nk}R_{nk}, \qquad n=1,...,N \ , \\
\sum_{n=1}^N \alpha_{nk}=1 , \qquad k=1,...,K. \ena \eeq The
Lagrangian is given by:
 \beq
 \bea{ll}
f\left(\vga, \vgd, \vgm, \vgl,c\right)=&-c-\sum_{n=1}^N\
\delta_n\left(\sum_{k=1}^K \alpha_{nk}R_{nk}-c/\grg_n
\right) \\
& -\sum_{n=1}^N\sum_{k=1}^K \mu_{nk}\alpha_{nk}\\
&+\sum_{k=1}^K\lambda_k\left(\sum_{n=1}^N \alpha_{nk}-1\right)-\beta
c.
 \ena
 \eeq
 To better understand the problem, we first derive the KKT conditions.
Taking the derivative with respect to the variables $\alpha_n(k)$
and $c$ we obtain
 \beq
  \label{KKT1}
\left\{
  \begin{array}{c}
- \mu_{nk}+\lambda_k-\delta_n R_{nk}=0 \\
-1+\sum_{n=1}^N\frac{\delta_n}{\grg_n}-\beta=0,
\end{array}
\right.
\eeq
with the complementarity conditions:
 \beq
 \left\{
 \bea{l}
 \label{KKT2}
 \gl_n\left(\sum_{n=1}^N\alpha_{nk}-1\right)=0, \\
 \delta_n\left(\sum_{k=1}^K \alpha_{nk}R_{nk}-c/\grg_n
 \right)=0,\\
 \mu_{nk}\alpha_{nk}=0, \\
 \beta c=0,\mu_{nk}\geq0, \beta\geq0, \delta_n\geq0.
 \ena
 \right.
 \eeq
 Note that this problem is always feasible by choosing $c=0$.
Based on (\ref{KKT1})-(\ref{KKT2}) we can easily see that the following proposition holds:
\begin{prop}
\label{KKT_eq}
The Lagrange multipliers in equation (\ref{KKT2}) satisfy the following claims:\\
\begin{itemize}
\item[1.] If there is a non zero feasible solution then $\beta=0$.
\item[2.] For each user with total rate equal to $c>0$,
$\delta_n>0$, and $\gb=0$. Therefore, $\sum_{n=1}^N \delta_n/\grg_n=1$. Otherwise, $\delta_n=0$.
\item[3.] If $\alpha_{nk}>0$, then $\mu_{nk}=0$ and
$\lambda_k=\delta_n R_{nk}$.
\item[4.] If $\alpha_{nk}=0$, then $\mu_{nk}\ge0$ and
$\lambda_k\ge \delta_n R_{nk}$.
\end{itemize}
\end{prop}
From these we obtain the following proposition:
\begin{prop}
 \label{wmaxmin} The weighted max-min fair solution is achieved
if all users have equal weighted rates; i.e., the optimal $c$
satisfies for all $n$ $c=\grg_n R_n$.
\end{prop}
{\bf Proof: }Let $c$ be the optimal value. Assume that there is a
user $n$ with a rate higher than $c$ and let $k$ be a frequency such
that $\alpha_{n}\left(k\right)>0$. Define $\ga'_n(k)=\ga_n(k)-\gre$,
and for $m \neq n$: $\ga'_m(k)=\ga_m(k)+\gre/(N-1)$. Obviously the
weighted rate for all other users is increased. Choosing $\gre \le
\grg_n \sum_{k=1}^K \ga_n(k)R_{nk}-c$, ensures that $R_n>c$. Since
by construction all users $m \neq n$ achieve a rate higher than $c$
we obtain a contradiction to the optimality of $c$. This claim is
important result from  a network planning perspective. The achieved
rates are proportional to $1/\grg_n$; in other words, users with
rates $\grg_m,\grg_n$ will receive rates satisfying
$R_m/R_n=\grg_m/\grg_n$. This is desirable since utility typically
scales with $\log R$, so that doubling the rate results in a fixed
increase in the total utility.
\subsection{Voice and data rate allocation}
In networks carrying mixed services, it is important to be able to allocate a fixed bandwidth, to constant-bit-rate and latency-sensitive services such as voice services. The weighted max-min formulation can be easily generalized to this case.
 Voice users (fixed rate) will get at least
$R_{min}$, while, other variable-bit-rate users will get the
weighted max-min rate according to their respective service levels.
We have two groups of users: $V,D$ and the optimization becomes:
\beq
\max_{\vga_1,...,\vga_N,c} c\\
\left\{
\bea{l}
 0 \le c \\
 c\leq \sum_{k=1}^K \ga_{ik}R_{nk}, \qquad i \in D\\
 R_{\min} \le \sum_{k=1}^K \ga_{ik}R_{nk}, \qquad i \in V \\
  \sum_{i=1}^N \ga_i(k)=1 , \qquad k=1,...,K. \ena
  \right.
\eeq Here, one should solve the optimization problem first assuming
that the set $D$ is empty. This will confirm that there is a
feasible solution for the voice users. If there is a feasible
solution for the set $V$ then we know that there is a feasible
solution to the general problem. A simple version of this scenario
is analyzed in Example II in section \ref{simulation}.

We now show that the feasibility of a given rate allocation can be tested by solving
a simple weighted max-min problem, where the weights are given by
the inverse of the desired rates. By proposition \ref{wmaxmin} the
solution to the weighted max-min problem with weights given by
$\grg_n=1/R_n^d$ where $R_n^d$ is
the desired rate for user $n$, provides the largest $c$ such that
for each user $c R_n^d=R_n$. Hence the rate vector
$\left(R_1^d,...,R_N^d\right)$ is feasible if and
only if the solution satisfies $1 \le c$. Otherwise the rate vector
is infeasible. This completes the solution of the feasibility problem. Note that the solution holds even when each constant bit-rate user has a different rate requirement.
\section{The two user case}
In this section, we address the special cases of two users. In this
case the optimization problem can be dramatically simplified.
 Using $1-4$ in
proposition \ref{KKT_eq}  above we can easily conclude
 that the partition rules are as follows:
\begin{enumerate}
\item $\frac {\delta_1}{\grg_1}+\frac{\delta_2}{\grg_2}=1$. Special
case of item 2 in proposition  \ref{KKT_eq}.
\item If $\delta_1 R_{1k}>\delta_2 R_{2k}$ the frequency bin $k$ is
allocated to user 1.
\item If $\delta_1 R_{1k}<\delta_2 R_{2k}$ the frequency bin $k$ is
allocated to user 2.
\item If $\delta_1 R_{1k}=\delta_2 R_{2k}$ the frequency bin $k$ is
shared between the users such that they both get the same total
rate. based on item 3 in proposition  \ref{KKT_eq}.
\end{enumerate}
 An interesting
consequence of our analysis is that in the two user case at most a
single subcarrier should be shared between the users. This
conclusion can be extended to the $N$ user case, where at most $N
\choose 2$ frequencies are shared in time. The proof is given in
appendix I.

Based on the above properties we suggest an $O(K \log_2 K)$
complexity algorithm motivated by our analysis of the Nash
Bargaining Solution (NBS) for the frequency selective interference
channel \cite{leshem2006}, \cite{leshem2008}. Extensions to the
total power constraint are possible, similarly to the solution of
the NBS \cite{leshem2009gamenets}. We also show that at most a
single frequency may be shared between the two users. To that end,
let $\alpha_{1k}=\alpha_k$, and $\alpha_{2k}=1-\alpha_k$, and
without loss of generality, we set  $\grg_{1}=1$, and
$\grg_{2}=\grg$. The ratio
$\Gamma=\frac{\delta_2}{\delta_1}=\frac{1-\delta_1}{\delta_1\grg}$
is a threshold which is independent of the frequency and is set by
the optimal assignment. Although $\Gamma$ is a-priori unknown, it
exists. We also assume that the rate ratios $L(k)=R_{1k}/R_{2k},
1\leq k\leq K$ are sorted in decreasing order; i.e. $L(k)\geq L(k'),
\forall k\leq k'.$ \footnote{This can be achieved by sorting the
frequencies according to $L(k)$.} Using proposition \ref{wmaxmin} we
obtain
 \beq \sum_{k=1}^K
\ga_kR_{1k}=\grg\sum_{k=1}^K \left(1- \ga_k\right)R_{2k}.
 \eeq
We are now ready to define the optimal assignment of the
$\alpha_k$'s.

Let $\Gamma_k$ be a moving threshold defined by \beq
\Gamma_k=\frac{A_k}{B_k \grg} \eeq where \beq A_k=\sum_{m=1}^k
R_{1m}, \ B_k=\sum_{m=k+1}^K R_{2m}. \eeq $A_k$ is a monotonically
increasing sequence, while $B_k$ is monotonically decreasing. Hence,
$\Gamma_k$ is also monotonically increasing. $A_k$ is the rate of
user 1 respectively when frequencies $1,...,k$ are allocated to him.
Similarly $B_k$ is the rate of user 2 when frequencies $k+1,...,K$
are allocated to him. Let \beq \label{def_k}
k_{\min}=\min_k\left\{k: A_k \ge B_k \grg\right\}. \eeq We are
interested in a feasible solution such that the ratio of the
accumulated rate of the users will be equal to $\grg$. Thus,
frequency bin $k_{min}$ has to be split between the users, and
$\alpha_{k_{min}}$ is given by \ \beq \label{def_alpha}
A_{k_{min}-1}+\alpha_{k_{min}}R_{1k_{min}}=
\grg\left(B_{k_{min}-1}-\alpha_{k_{min}}R_{2k_{min}}\right), \eeq or
\beq \alpha_{k_{min}}=\frac{\grg
B_{k_{min}-1}-A_{k_{min}-1}}{R_{1k_{min}}+\grg R_{2k_{min}}}. \eeq
It easy to confirm that $0\le\alpha_{k_{min}}\le 1$.

The outline of the algorithm is given in Table
\ref{two_players_table}.

\begin{table}
\centering \caption{Algorithm for computing the 2x2 weighted
max-min}
\begin{tabular}{||l||}
\hline \hline {\bf Initialization:}  Sort the ratios $L(k)$ in decreasing order. \\
Calculate the values of $A_k,B_k$ and $\Gamma_k$. \\
\hline
Calculate $k_{min}$ using (\ref{def_k}). \\
Calculate $\ga_{k_{min}}$ using (\ref{def_alpha}). \\
User $1$ gets the bins $1:k_{min-1}$ and $\ga_{k_{min}}$ of bin $k_{min}$. \\
User $2$ gets the bins $k_{min+1}:K$ and $1-\ga_{k_{min}}$ of bin $k_{min}$.\\
\hline \hline
\end{tabular}
\label{two_players_table}
\end{table}

\begin{table}
\centering
\begin{tabular}{|c|c|c|c|c|c|c|}

  \hline
 $k$                 & 1 & 2 & 3 & 4  & 5 & 6  \\
 \hline
 $R_1$            & 14 & 18 & 5 & 10 & 9 & 3 \\
  \hline
 $R_2$            & 6 & 10 & 5 &15 & 17&16 \\
  \hline
 $L\left(k\right)$ &2.33 &1.80 &1.00& 0.67& 0.53 &0.19 \\
  \hline
 $A_k$ & 14 & 32 & 37 &47 &56 &59 \\
   \hline
 $B_k$ & 63 & 53& 48& 33& 16 &0 \\
 \hline
 $\Gamma_{k}$ & .178& .483 &.617 &1.14 &2.80 &$\infty$ \\
\hline
\end{tabular}
\caption{User rates in each frequency bin after sorting, and the
values of $\Gamma_k$.}
  \label{sorttable}
\end{table}

\section{Examples and Simulations}
\label{simulation} In this section we  report simulation results on
rate allocation for various values of weights.

\begin{figure}
  \begin{center}
    \mbox{\psfig{figure=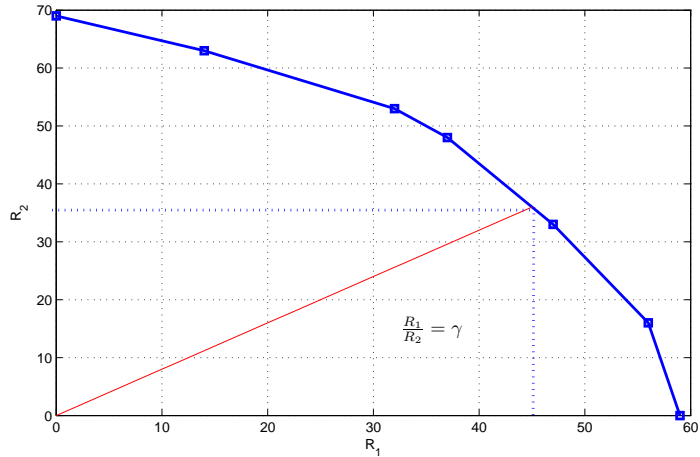,width=0.6\textwidth}}
 \end{center}
  \caption{The feasible total rate of player $1$ vs, the feasible
total rate of player $2$.}
  \label{FDM_rate_region}
 \end{figure}

To illustrate the algorithm we compute the weighted max-min solution
for the following example: {\em Example I}: Consider two users
communicating over a 2x2 memoryless Gaussian interference channel
with 6 frequency bins. The weights of user $1$ and $2$ are $1$ and
$1.25$, respectively. The interference free user rates in each
frequency bin (sorted according to $L_k$) are given in Table
\ref{sorttable}. We now compute the values of  $A_k$ and $B_k$ for
each user.  Since, $\Gamma_{3}>1$ we conclude that $k_{min}=4$ and
$\ga_{k_{min}}=0.8$.  Thus, user $1$ is using subcarriers $1,2,3$,
and sharing subcarrier $4$ with user $2$. The total rate of players
$1$ and $2$ are $45$ and $36 $, respectively.
 We can also give a geometrical interpretation to the solution. In Figure \ref{FDM_rate_region}
 we draw the feasible total rate that player $1$ can obtain as a function of the
total rate of player $2$. The enclosed area in blue, is the
achievable rates set. Since, the subcarriers are sorted according to
$L_k$ the set is convex. The point $\left(45,36\right)$ is the
operating point of the weight max-min with $\gamma=1.25$. A change
in the value of $\gamma$ will move the solution on the boundaries of
achievable rates set.

\begin{figure}
  \begin{center}
    \mbox{\psfig{figure=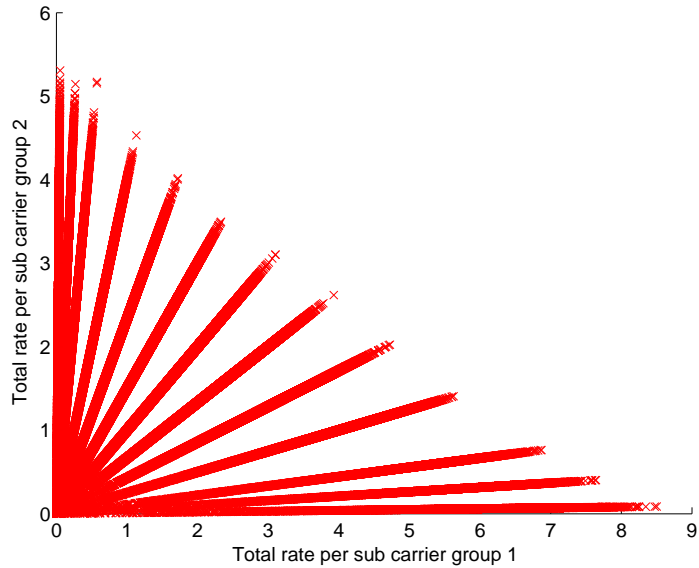,width=0.6\textwidth}}
 \end{center}
  \caption{The distribution of feasible rates for each value of $\gamma$. $[SNR_1,SNR_2] = [20dB,10dB]$.}
  \label{rays}
 \end{figure}

\begin{figure}
  \begin{center}
    \mbox{\psfig{figure=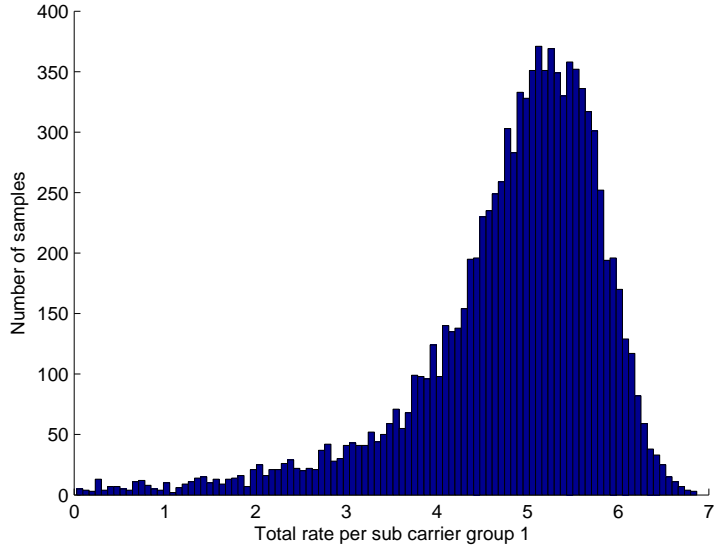,width=0.6\textwidth}}
 \end{center}
  \caption{A histogram of group 1 rates for $\gamma = 0.1$ and $[SNR_1,SNR_2] = [20dB,10dB]$.}
  \label{hist}
 \end{figure}

\begin{figure}
\begin{center}
\mbox{\psfig{figure=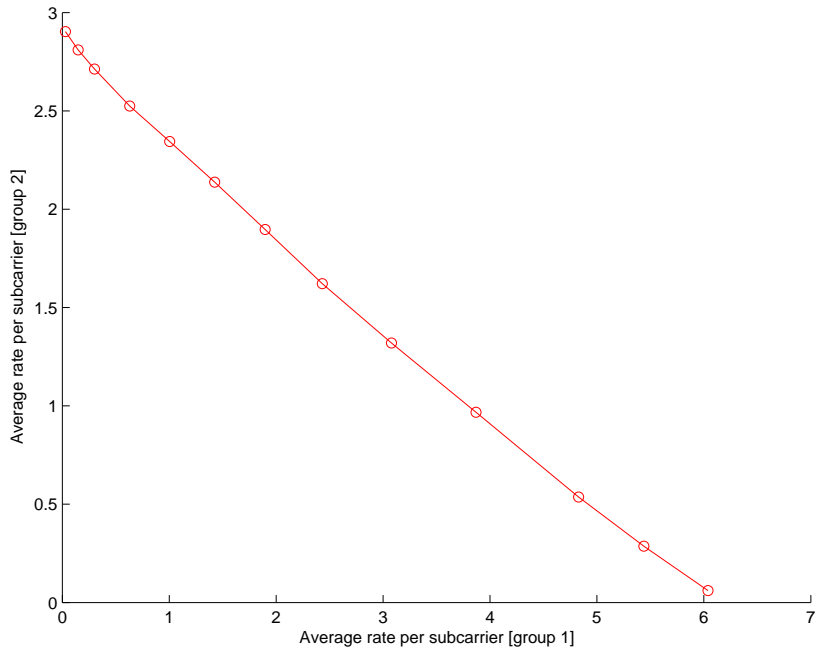,width=.6\textwidth}}
\end{center} \caption{The average rate of group $2$ vs. the average rate of
group $1$ for $[SNR_1,SNR_2] = [20dB,10dB]$.}
\label{Avg_rate1}
\end{figure}

\begin{figure}
\begin{center}
\mbox{\psfig{figure=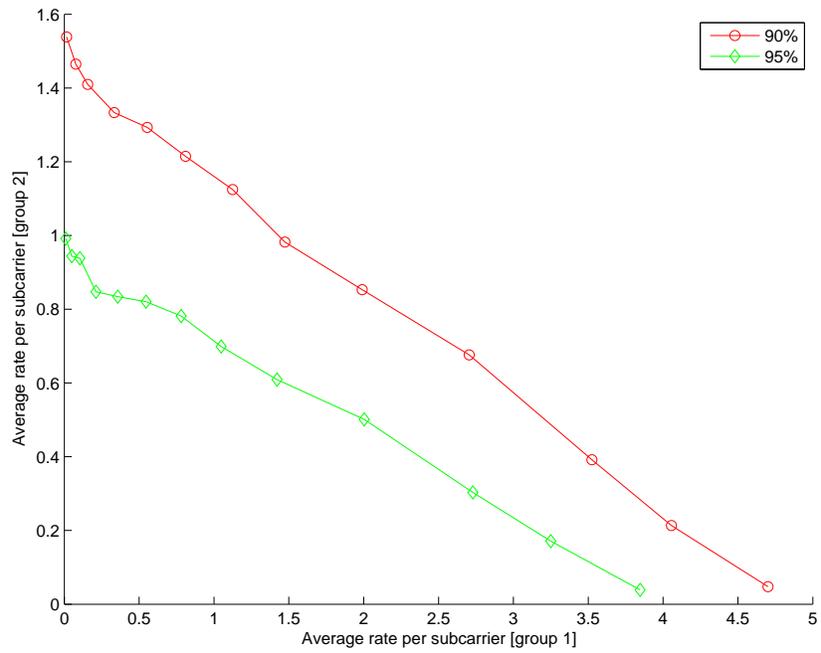,width=.6\textwidth}}
\end{center} \caption{The rate of group $2$ vs. the rate of group
$1$ for outage probabilities of $10$\% and $5$\%.  $[SNR_1,SNR_2]
= [20dB,10dB]$.} \label{success1}
\end{figure}

Next, we demonstrate  simulation results of rate allocation for various values of weights in two cases.
In both cases the users are communicating over a frequency selective Rayleigh fading channel with variance 1.
The number of frequency bins is 64
\emph{Case 1, simulation of two data groups:}
The first case simulates two groups of users, each group is of size 8.  This is a typical scenario where
one group has higher priority. The weight for one data group
is $\gamma$ while for the second data group is $1-\gamma$, where $0 \le \gamma \le 1$.  For each value of $\gamma$ we have performed $10000$ tests. The $SNR$ values of the two data groups are  $20$dB and $10$ dB respectively.
Figure \ref{rays} presents the distribution of the feasible rates
for various value of $\gamma$. It is clear that for a given value of
$\gamma$ the feasible rate will be along a ray with an angle
$\phi=\arctan\frac{\gamma}{1-\gamma}$ relative to the $x$ axis. Figure \ref{hist} presents a histogram of the ray with $\gamma = 0.1$.
Figure \ref{Avg_rate1}, presents the average value of the feasible
rate for group $1$ vs. average rate of group $2$. Figure \ref{success1} shows the outage regions for outage
probability of $0.1$ and $0.05$.
We can clearly see that reducing the outage has significant impact on the achievable rates.

\begin{figure}
\begin{center}
\mbox{\psfig{figure=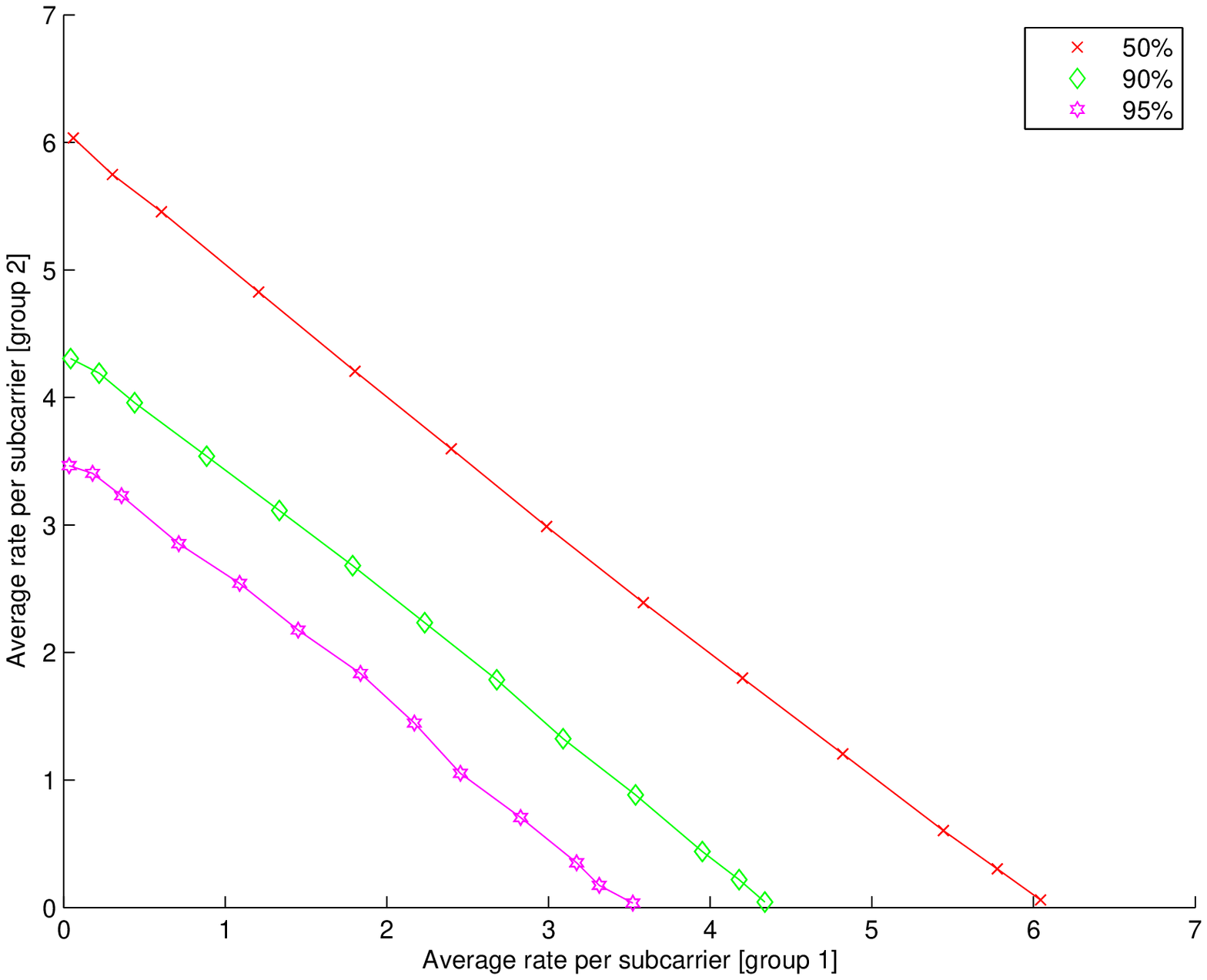, width=.6\textwidth}} \end{center}
\caption{The rate of data group $2$ vs. the rate of data group $1$ for
 $SNR= 20$ (voice group $SNR =5$), and outage probabilities $0.05$,$0.1$ and $0.5$.}
\label{case2success}
\end{figure}

\emph{Case 2, simulation of a voice group and two data groups:}
The second case simulates three groups of users, a voice group of size 4 and two data groups each of size 8.
The $SNR$ value of the voice group is $5$dB and the $SNR$ of the two data groups is $20$dB.  Figure \ref{case2success} shows the outage regions for outage probability of $0.05$,$0.1$ and $0.5$.


\section{Conclusion and extensions}
In this paper we described a simple rate allocation technique for
multiple-access OFDMA systems applying joint TDM/FDM subchannel
allocation. The method is applicable whenever a central access point
or base station is available. The complexity of the technique is
very low. Furthermore, the allocation can be done using channel
statistics instead of the actual channels. We have also demonstrated
how to accommodate and test the feasibility of a set of constant
rate users. Finally, we have analyzed the two user case, and
provided a very low complexity weighted max-min algorithm for this
case.

\section{Appendix}
{\bf Lemma I.1 :} Assume that all the rate ratios $R_1(k)/R_2(k)$
are different from each other then at most a single frequency bin is
shared between the two users.

 {\bf Proof: } Based on 3 in proposition \ref{KKT_eq}  a subcarrier is shared between two users if
 $\delta_1 R_{1k}=\delta_2 R_{2k}$, or in other words
 $\frac{\delta_2}{\delta_1}=\frac{R_{1k}}{R_{2k}}$. Hence, if all
 rate ratios are different, at most a single frequency may have a rate ratio equal to  $\frac{\delta_2}{\delta_1}$.

{\bf Lemma I.2 :} Assume that there is a solution where two
subcarriers are shared between the users. Then there is an alternative
solution where only a single subcarrier is shared between the users.

 {\bf Proof: }Assume without loss of generality that subcarriers  $1$, and $2$ are shared between users $1$ and $2$.
  User $1$ gets fractions $\alpha_1$ and
 $\alpha_2$ from subcarriers $1$ and $2$, respectively. User $2$ gets fractions $\beta_1$ and
 $\beta_2$ from subcarriers $1$ and $2$, respectively (where $\ga_i+\gb_i=1$).  Based on proposition \ref{KKT_eq} the rate ratios in these frequency bins should satisfy the relation
 $\frac {R_{11}}{R_{21}}=\frac {R_{12}}{R_{22}}$, and the total
 rate of each user satisfies the conditions:
 \beq
 \begin{array}{rcl}
 \frac{c}{\gamma_1}&=&A+\alpha_1 R_{11}+\alpha_2 R_{12}\\
 \frac{c}{\gamma_2}&=&B+\beta_1 R_{21}+\beta_2 R_{22}
 \end{array},
 \eeq
 where $A$ and $B$ are the sum of  rates of users $1$ and $2$ on the other  frequency bins.
 We note that in one hand, if $\alpha_1\frac{R_{11}}{R_{12}}\leq \beta_2$, then user $1$ can set $\alpha_1$ to $0$ while
 increasing his share in subcarrier $2$ by $\alpha_1\frac{R_{11}}{R_{12}}$. On the other hand, when $\alpha_1 \frac{ R_{11}}{R_{12}}> \beta_2$ we obtain
  $\alpha_1 > \beta_2 \frac{ R_{22}}{R_{21}}$. Therefore, user $2$ can set $\beta_2$ to $0$ and increase his  fraction in subcarrier $1$
  by $\beta_2\frac{ R_{22}}{R_{12}}$.

{\bf Lemma I.3 :} In the $N$ user case at most $N \choose 2$
frequencies are shared in time.

{\bf Proof} Based on Lemma I.2. at most a single frequency bin is shared
between any two users. Since the number of different pair of users
is  $N \choose 2$, then the maximum number of frequency bins that
are time shared is upper bounded by $N \choose 2$.

\bibliographystyle{ieeetr}


\begin{thebibliography}{10}

\bibitem{jang03}
J.~Jang and K.~B. Lee, ``Transmit power adaptation for multiuser {OFDM}
  systems,'' {\em IEEE Journal on Selected Areas in Communications}, vol.~21,
  pp.~171--178, Feb. 2003.

\bibitem{iyengar2006}
R.~Iyengar, K.~Kar, and B.~Sikdar, ``Scheduling algorithms for
  point-to-multipoint operation in {IEEE} 802.16 networks,'' in {\em 4th
  International Symposium on Modeling and Optimization in Mobile, Ad Hoc and
  Wireless Networks, 2006}, pp.~1--7, April 2006.

\bibitem{rhee2000}
W.~Rhee and J.~Cioffi, ``Increase in capacity of multiuser ofdm system using
  dynamic subchannel allocation,'' in {\em IEEE 51st Vehicular Technology
  Conference Proceedings, 2000. VTC 2000-Spring Tokyo. 2000}, vol.~2, pp.~1085
  --1089 vol.2, 2000.

\bibitem{shen05}
Z.~Shen, J.~G. Andrews, and B.~L. Evans, ``Adaptive resource allocation in
  multiuser {OFDM} systems with proportional rate constraints,'' {\em IEEE
  Transactions on Wireless Communications}, vol.~4, no.~6, pp.~2726--2737,
  2005.

\bibitem{han2005}
Z.~Han, Z.~Ji, and K.~Liu, ``Fair multiuser channel allocation for {OFDMA}
  networks using the {N}ash bargaining solutions and coalitions,'' {\em IEEE
  Trans. on Communications}, vol.~53, pp.~1366--1376, Aug. 2005.

\bibitem{nokleby07}
M.~Nokleby, A.~Swindlehurst, Y.~Rong, and Y.~Hua, ``Cooperative power
  scheduling for wireless {MIMO} networks,'' {\em IEEE Global
  Telecommunications Conference, 2007. GLOBECOM '07.}, pp.~2982--2986, Nov.
  2007.

\bibitem{leshem2009gamenets}
E.~Zehavi and A.~Leshem, ``Bargaining over the interference channel with total
  power constraints,'' in {\em International Conference on Game Theory for
  Networks, 2009. GameNets '09.}, pp.~447 --451, 13-15 2009.

\bibitem{leshem2008}
A.~Leshem and E.~Zehavi, ``Cooperative game theory and the {G}aussian
  interference channel,'' {\em IEEE Journal on Selected Areas in
  Communications}, vol.~26, pp.~1078--1088, September 2008.

\bibitem{park07}
H.~Park and M.~van~der Schaar, ``Bargaining strategies for networked multimedia
  resource management,'' {\em IEEE Transactions on Signal Processing}, vol.~55,
  pp.~3496 --3511, july 2007.

\bibitem{zehavi09b}
E.~Zehavi and A.~Leshem, ``Alternative bargaining solutions for the
  interference channel,'' in {\em 3rd IEEE International Workshop on
  Computational Advances in Multi-Sensor Adaptive Processing (CAMSAP), 2009},
  pp.~9 --12, 13-16 2009.

\bibitem{chen09}
J.~Chen and A.~Swindlehurst, ``Downlink resource allocation for multi-user
  mimo-ofdma systems: The kalai-smorodinsky bargaining approach,'' in {\em 3rd
  IEEE International Workshop on Computational Advances in Multi-Sensor
  Adaptive Processing (CAMSAP), 2009}, pp.~380 --383, 13-16 2009.

\bibitem{biglieri98}
E.~Biglieri, J.~Proakis, and S.~Shamai, ``Fading channels:
  information-theoretic and communications aspects,'' {\em IEEE Transactions on
  Information Theory}, vol.~44, pp.~2619 --2692, oct 1998.

\bibitem{leshem2006}
A.~Leshem and E.~Zehavi, ``Bargaining over the interference channel,'' in {\em
  Proc. IEEE ISIT}, 2006.

\end{thebibliography}

\end{document}